# Dependence of Physiochemical Features on Marine Chlorophyll Analysis with Learning Techniques

Subhrangshu Adhikary[1], Sudhir Kumar Chaturvedi[2,*], Saikat Banerjee[3,] Sourav Basu[4]

[1]Department of Computer Science and Engineering, Dr. B.C. Roy Engineering College, Durgapur-713206, West Bengal, India
subhrangshu.adhikary@spiraldevs.com
[*,2]Department of Aerospace Engineering, UPES, Dehradun-248007, India
sudhir.chaturvedi@ddn.upes.ac.in
[3]Department of Mechanical engineering, Cubicx, Kolkata-700070, West Bengal, India
*saikatbanerjee@cubicxindia.com*
[4]Department of Electrical engineering, Cubicx, Kolkata-700070, West Bengal, India
*souravbasu@cubicxindia.com*

**Abstract.** Marine chlorophyll which is present within phytoplankton are the basis of photosynthesis and they have a high significance in sustaining ecological balance as they highly contribute toward global primary productivity and comes under the food chain of many marine organisms. Imbalance in the concentrations of phytoplankton can disrupt the ecological balance. The growth of phytoplankton depends upon the optimum concentrations of physiochemical constituents like iron, nitrates, phosphates, pH level, salinity, etc. and deviations from an ideal concentration can affect the growth of phytoplankton which can ultimately disrupt the ecosystem at a large scale. Thus the analysis of such constituents has high significance to estimate the probable growth of marine phytoplankton. The advancements of remote sensing technologies have improved the scope to remotely study the physiochemical constituents on a global scale. The machine learning techniques have made it possible to predict the marine chlorophyll levels based on physiochemical properties and deep learning helped to do the same but in a more advanced manner simulating the working principle of a human brain. In this study, we have used machine learning and deep learning for the Bay of Bengal to establish a regression model of chlorophyll levels based on physiochemical features and discussed its reliability and performance for different regression models. This could help to estimate the amount of chlorophyll present in water bodies based on physiochemical features so we can plan early in case there arises a possibility of disruption in the ecosystem due to imbalance in marine phytoplankton.

## 1 Introduction

Phytoplanktons are marine organisms coexisting with other marine organisms and together they establish a balance of the ecosystem as they become part of the food chain for many other sea creatures as well as they are responsible for primary productivity and contributes significantly toward global oxygen production. Phytoplanktons are autotrophs which means they can prepare their food by using organic sources like carbon dioxide, water, etc. with the help of sunlight. The photosynthesis undergone by phytoplankton means that they have a pigment in their cell called chlorophyll. A chlorophyll can reflect near-infrared light bands by which satellites can remotely detect them [1]. Therefore remote sensing could be useful to remotely monitor phytoplankton blooming. Remote sensing could also be used to detect many other physiochemical features. EU Copernicus Program has used these remote sensing abilities to create a publicly available reanalysis data which we have used in this experiment. The basic remote sensing data are used in different combination to create reanalysis data which contains features like primary productivity, pH levels, salinity, phosphates, nitrates, etc.

The amount of phytoplankton in the ocean directly or indirectly depends upon the concentration of many physiochemical features and phytoplanktons require a balanced amount of these features for optimum growth. Deviation of these features could deplete the phytoplankton levels. Therefore reanalysis data of remote sensing can be useful to detect volumes of both chlorophyll and other physiochemical features, this leaves an immense opportunity to remotely monitor them. On the other hand, due to the advancement of machine learning and deep learning technologies, we can monitor the chlorophyll levels automatically utilizing remote sensing.

In this experiment, we have tried to establish a relationship between chlorophyll with physiochemical features utilizing different machine learning and deep learning techniques to detect chlorophyll levels based on these features remotely and automatically. We have chosen the Bay of Bengal as our study location as it holds lots of ecological importance because of the Sundarbans Delta, which is home for many rare animals and sea creatures that co-exist.

The imbalance of the ecosystem for either of land or sea could affect one another directly or indirectly. The coordinates of the study location are from 5° N to 15° N and 82° E to 92° E.

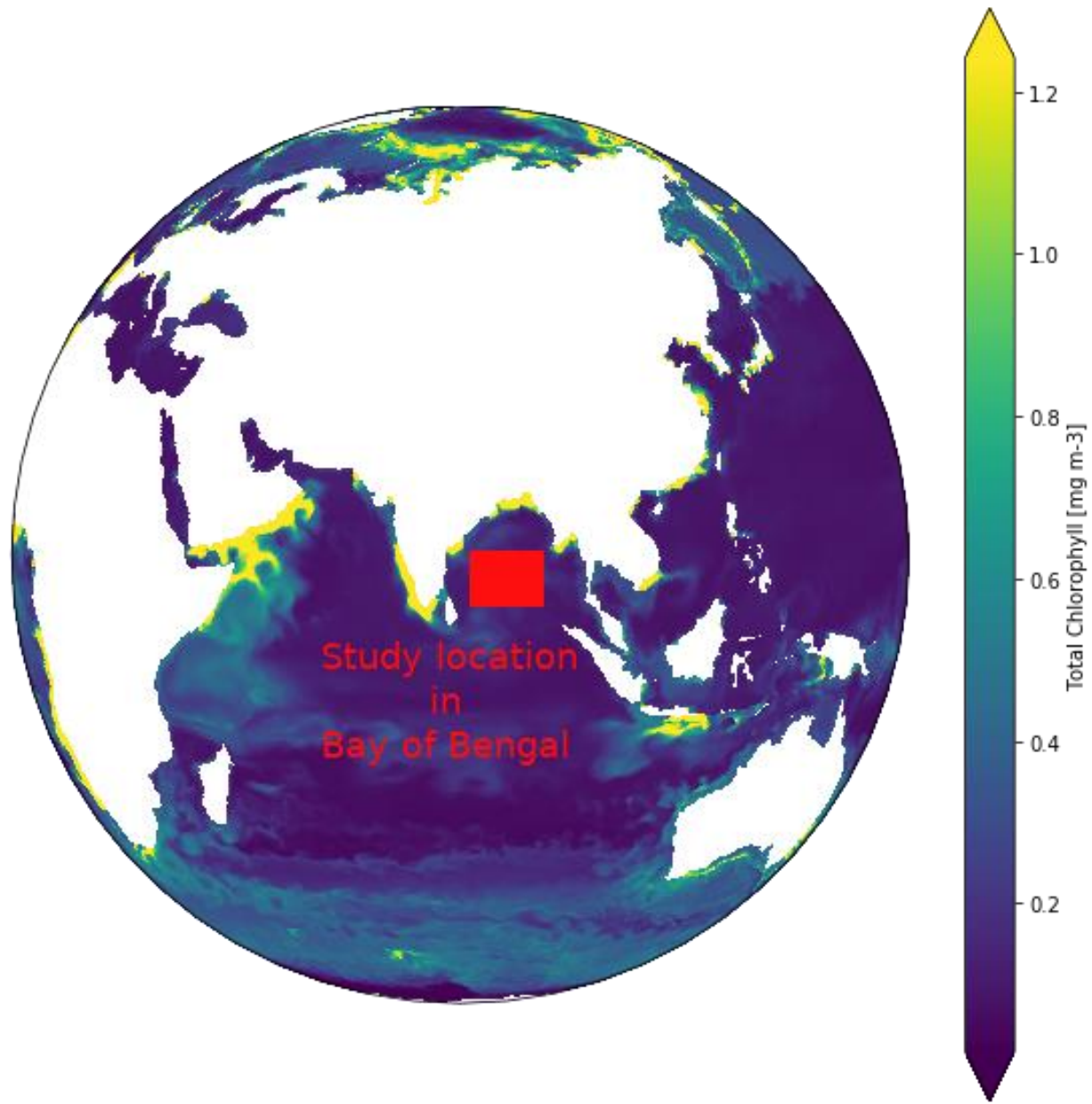

**Fig. 1.** A map of chlorophyll distribution at depth 0.5 *m* of sea-level indicating our study location in the Bay of Bengal

## 2 Recent Advancements in Marine Chlorophyll Study

Here in this section, we will discuss the advancements in the field of chlorophyll study, remote sensing, machine learning and deep learning studies and will discuss why we have chosen this study area.

Chlorophylls are present within phytoplankton which makes it capable of performing photosynthesis. Chlorophyll present within phytoplanktons could be detected by phaeophytin by fluorescence method [2]. Earlier, N, N-dimethylformamide has been introduced as an upgraded method for phytoplankton chlorophyll detection [3]. The phytoplankton distribution in the southern ocean was studied to understand the primary production capacity of marine phytoplanktons [4]. A model was prepared in [5] for the carbon ratio and the conversion factor between productivity and growth rate of phytoplankton. The thermal structure of the ocean and its consequences of non-uniformity in chlorophyll profile was discussed in [6]. The nutrient requirements and importance of nitrogen requirements as well as its storage in macroalgae was established [7].

Earlier, an early-warning protocol was developed for prediction of chlorophyll-a concentrations with machine learning for the estuarine reservoirs, Korea with a 7-year monitoring data [8]. Geostationary Ocean Color Imager satellite data was earlier used to monitor coastal water quality with help of machine learning technique and obtained low accuracy of 1.74 $mg\ m^{-3}$ root mean squared error [9]. The application of machine-learning techniques was used to create a consistent and calibrated global chlorophyll concentration baseline dataset by utilizing remote sensing data for ocean colour [10].

Random Forest Classifier was used for classification of remote sensing data obtained from Landsat Enhanced Thematic Mapper Plus (ETM+) and its performance was compared with support vector machine classifier [11]. The same classifier along with Decision Tree Classifier was used for assessment of chlorophyll sufficiency levels of mature palm oil utilizing hyperspectral remote sensing technology [12]. Decision tree was also used to classify transparent plastic-mulched landcover utilizing Landsat-5 TM images [13]. Artificial neural network-based approach with the help of multi-layer perceptron to predict leaf chlorophyll content by analyzing cotton plant images [14]. MLP was also proven to be effective for estimation of chlorophyll concentration index of lettuce and obtained coefficient of determination 0.98 and a mean squared error 0.006 during the validation process. Convolution neural network was used for the estimation of seawater chlorophyll-a by analyzing hyperspectral images [15]. Image-based canopy reflectance model was developed for remote estimation of one-sided leaf area and leaf chlorophyll content by utilizing CNN technologies [16]. Deep neural network with the help of CNN was also used for chlorophyll-a concentration model with an extremely imbalanced dataset and showed that proper pre-processing techniques and oversampling can improve prediction accuracy in these cases [17]. Deep learning was used for investigation of chlorophyll-a and total suspended matter concentration in Taihu Lake, China utilizing Landsat ETM and field spectral measurements [18].

## 3 Methodology
Here in this section, we have talked about the process of data collection, preprocessing and classification of the data in details.

### 3.1 Data Availability and Preprocessing
EU Marine Copernicus Program is a widely accepted platform for accessing publicly available remote sensing data. They have used basic satellite remote sensing data to create a reanalysis data which indicates the concentrations of different physiochemical items like surface carbon dioxide pressure, dissolved oxygen, chlorophyll, dissolved silicates, pH levels, total primary productivity, etc. at a depth of 0.5 $m$. We have downloaded these data for the year 1993 to 2018 in NetCDF format. This data contains the historical data in the form of time series for each location coordinates at 4 $km$ pixel resolution. Now, we have converted this file to a time series statistical table for each location totalling up to over thirty-seven thousand rows to train on. To indicate low levels of chlorophyll, we have marked 1/4[th] of the dataset having the lowest amount of chlorophyll as "Critical" and marked remaining as "Normal". Later we have randomly split up the table into training and testing set containing 75% and 25% data respectively. The summary of the time series data is mentioned in table 1 and a portion of the time series is represented by fig. 2.

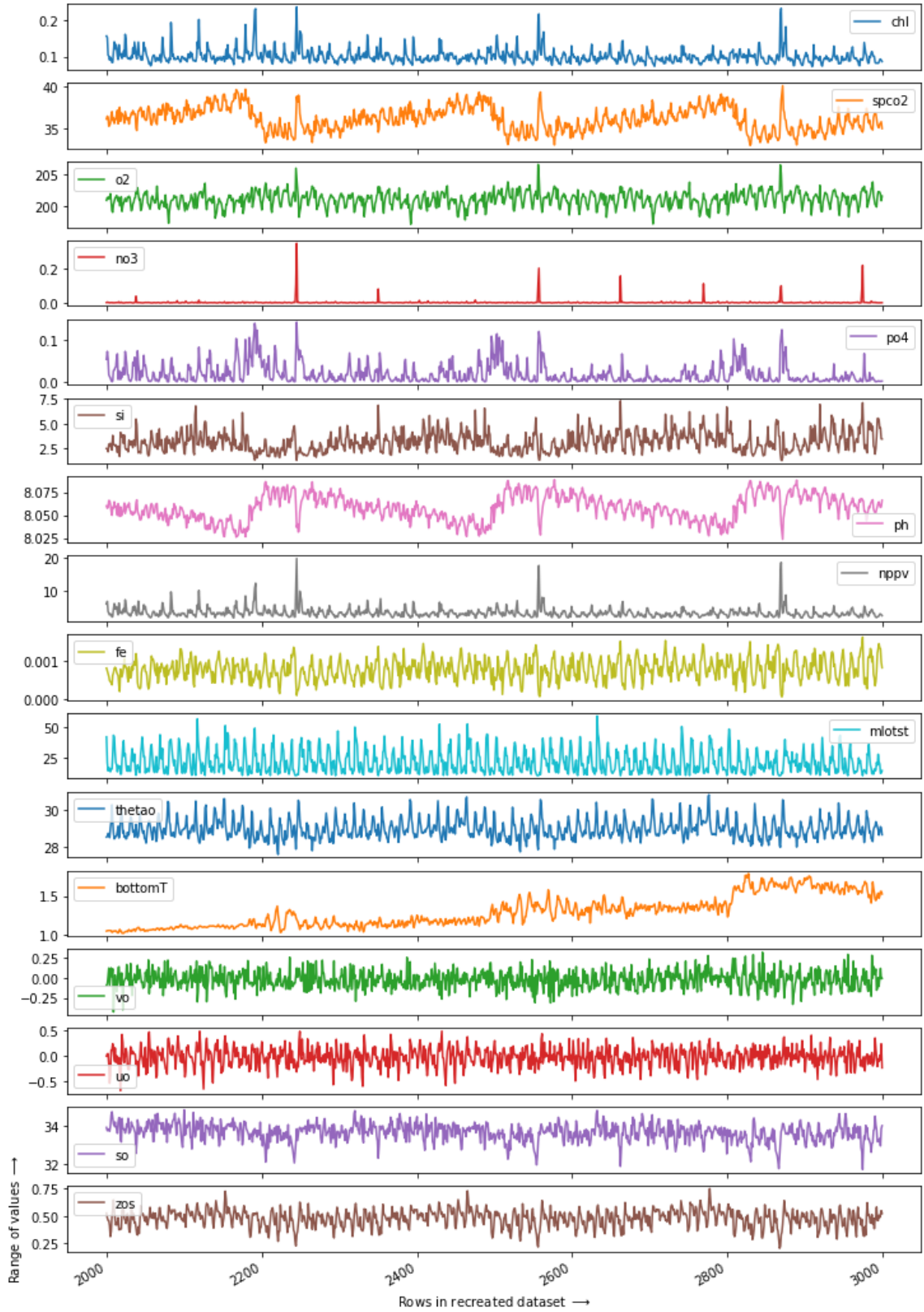

**Fig. 2.** A graph of all the features from a small portion of the entire dataset as a sample.

**Table 1.** The summary of the dataset

| Feature | Min | Max | Mean | Standard Deviation | Unit |
|---|---|---|---|---|---|
| spco2 | 28.3267 | 41.4043 | 35.877050 | 1.647791 | $Pa$ |
| $O_2$ | 194.291 | 219.503 | 202.106733 | 2.703826 | $mmol\ m^{-3}$ |
| Chl | 0.0726752 | 1.31307 | 0.120117 | 0.058221 | $mg\ m^{-3}$ |
| $NO_3$ | 0.00155384 | 7.94053 | 10.000000 | 0.419270 | $mmol\ m^{-3}$ |
| $PO_4$ | 4.00281e-05 | 0.414416 | 0.022617 | 0.031002 | $mmol\ m^{-3}$ |
| si | 0.720682 | 11.9874 | 3.283496 | 1.289600 | $mmol\ m^{-3}$ |
| pH | 8.01069 | 8.13985 | 8.059485 | 0.016567 | - |
| nppv | 1.21979 | 82.9279 | 4.686437 | 4.723566 | $mg\ m^{-3}\ day^{-1}$ |
| Fe | 3.91981e-05 | 0.00294045 | 0.001006 | 0.000363 | $mmol\ m^{-3}$ |
| mlotst | 10.5289 | 66.2252 | 20.445362 | 7.716653 | $m$ |
| thetao | 25.4276 | 31.463 | 28.749488 | 0.846414 | °C |
| bottomT | 0.654163 | 20.5935 | 1.620214 | 1.627649 | °C |
| vo | -1.5656 | 1.20792 | 0.007621 | 0.165873 | $m\ s^{-1}$ |
| uo | -0.913724 | 1.37638 | -0.000650 | 0.218761 | $m\ s^{-1}$ |
| so | 28.0969 | 35.4183 | 33.248507 | 0.702603 | $10^{-3}$ |
| zos | 0.146489 | 0.979339 | 0.505828 | 0.088441 | $m$ |

## 3.2 Classifiers

**Random Forest Classifier (RFC)**

Random Forest Classifiers are powerful ensemble-based classifier which creates several trees to make a decision with each and then the output of the classifier becomes the class which have received the majority of these trees' decisions. To decide the number of nodes of the trees to make a decision, the Gini Index or Entropy function is used. We have used the Gini Index method for this experiment to create branches of the trees. As RFC classifier works by making several trees and reach a decision based on their output, it becomes very slow compared to other classifiers for the purpose but its decisions become very accurate and stable.

**Decision Tree Classifier (DTC)**

A decision tree is a supervised machine learning algorithm based on ensemble classification method. The features of the dataset are represented as nodes of a DTC and the branches of the tree represent a decision that follows and leaf nodes of the tree represent outcomes. Similar to RFC, DTC also uses the Gini Index and Entropy function method to decide nodes of the tree. A decision tree learns from the pathway of the decision made during the training process, therefore DTC often seems to overfit the model, which means it attempts to fit as many training data as possible and if the values of the testing set are too much different from the training set, DTC might start showing erroneous results. Therefore for an ideal DTC, the training data should have a close match for a testing data and in such situation, there needs to have enough amount of training data so a model could properly recognize close matches while predicting outputs.

**Multi Layer Perceptron (MLP)**

A perceptron is an artificial neuron and works similar to how a natural neuron works. They get signal from one end, it is modified and reach the other end from where it is transmitted to other perceptrons. An artificial neuron gets an input, multiply with some weight, add some bias and pass the result to next artificial neuron. When this process is repeated over a range of neurons forming stacks of different layers to create a network, it is called a Multi Layer Perceptron and it is the simplest form of neural network. The weights and biases of each node are first initialized with random values and then the cost of the network is monitored to tune the weights and biases to minimize loss. The goal is to create a path in the network passing through which inputs will mimic the ideal output. Dropout functions are often used to introduce randomness in the network by randomly activating and deactivating different nodes in the network to make the model learn and not memorize. It is a very powerful technique which mimics the decision making process of a human brain.

**Convolution Neural Network (CNN)**

Convolution neural network is an extension of MLP which primarily specializes to deal with image data however they are very well suited for statistical classification as well. A CNN first extracts several features from the dataset like grey level cooccurrence matrix, different edge detection techniques, k-means clustering and several more. However all extracted data are not necessarily useful therefore a max-pooling concept is used which extracts the most promising features from the feature extraction matrix. This discards unnecessary features and only keeps important ones so the model does not consume too many resources in the process. After a convolution layer, MLP is added to the base of it to make it more suitable for classification, this includes adding several hidden dense layers, dropouts, and activation functions which modifies the output of a node or a layer based on given rules. Finally, the output of the CNN network is used to make the decision.

## 4 Results and Discussion

Here in this section, we will discuss the performance of our models in two stages. First, we will use all available features to train our model and second we will use only the top 4 features which influence classification the most and we will discuss them in details.

**4.1 Classification with all available features**
When used all the features available in our dataset, we have observed that RFC obtained the highest classification accuracy of 93.92% followed by CNN, DTC and MLP obtaining 92.62%, 91.52% and 90.78% classification accuracy respectively. Now, accuracy is a metric which is very useful when using a balanced dataset for training means both the classes in the classifier should have an equal number of items. It's like if the dataset has 90 items of one class and 10 items of other class and our model labels all of them as the first class, it will still show 90% accuracy while it cannot identify any of the items from other class. Our dataset is unbalanced as it's 1/4$^{th}$ data are labelled as "Critical" and other 3/4$^{th}$ data are labelled "Normal", therefore we need more metrics to decide how our model performed. Two such important metrics are precision and recall. Precision is the ability of the classifier not to detect a false-positive item and recall is the ability of the classifier not to detect a false-negative item in the dataset. Closer the score is to 1, better the model performs.

We can see that RFC has obtained a precision score of 0.923 which is quite impressive, and CNN received 0.916. However, both DTC and MLP have a lower precision score which indicates that they have labelled many items as false-positive. This indicates that RFC and CNN have better prediction than DTC and MLP in terms of precision. Now when we consider recall, we can see that RFC has the highest recall but DTC, MLP and CNN have a lower recall score indicating that they have labelled many items as a false negative. As far as accuracy, precision and recall are concerned, RFC stands out to be the best classifier for the situation.

Now when it comes about resource optimization, we can see that DTC was the fastest to train by training the model in 400.63 *ms*, followed by RFC, MLP and CNN with 7878.39 *ms*, 11351.09 *ms* and 18057.22 *ms* respectively. For generating output, we can see that DTC again becomes the fastest to predict results in 1.668 *ms*, followed by MLP, RFC and CNN which took 5.22 *ms*, 139.494 *ms* and 207.026 *ms* respectively. From this, we can see that DTC by far is the fastest model for both training and testing. However, given its low precision and recall score, we choose RFC as an alternative when we need fast training. But RFC fails to generate results fast, therefore looking at the testing time, MLP ranked second. However, accuracy, precision and recall of MLP are lower than DTC and therefore it cannot be fully reliable either. Therefore, going by the third in the ranking, although it performs slower than the other classifiers, we conclude RFC to be the best classifier based on overall performance, and when speed is the ultimate goal, DTC could also be alternatively used by compromising around 2.4% accuracy. The performances are recorded in table 2.

Table 2. Classification performance while using all features

| Classifier | Training Time (ms) | Testing Time (ms) | Accuracy (%) | Precision | Recall |
|---|---|---|---|---|---|
| Random Forest | 7878.39 | 139.494 | 93.92 | 0.923 | 0.917 |
| Decision Tree | 400.63 | 1.668 | 91.52 | 0.884 | 0.89 |
| Multi Layer Perceptron | 11351.09 | 5.22 | 90.78 | 0.882 | 0.875 |
| Convolution Neural Network | 18057.22 | 207.026 | 92.62 | 0.916 | 0.894 |

## 4.2 Using only the best 4 features to increase model speed

Using each of features available in our dataset to train the model, although increases overall accuracy, reduces the model building time and prediction time. Therefore to reduce this time, we can use fewer but more relevant features which although reduces training and testing time, it trades off some of the accuracies in the process. Our goal is to minimize this accuracy loss and maximize resource optimization. For this purpose, we have used the pair plot technique to find the most relevant features for our training. By pair plotting all the features against one another, we can observe that more separate the clusters of classes are, better it works as a classification feature. Fig. 3 shows the pair plot for 4 features whose clustered classes were most distinguishable among the rest of the models. Therefore by using them, we can expect a reasonable classification performance.

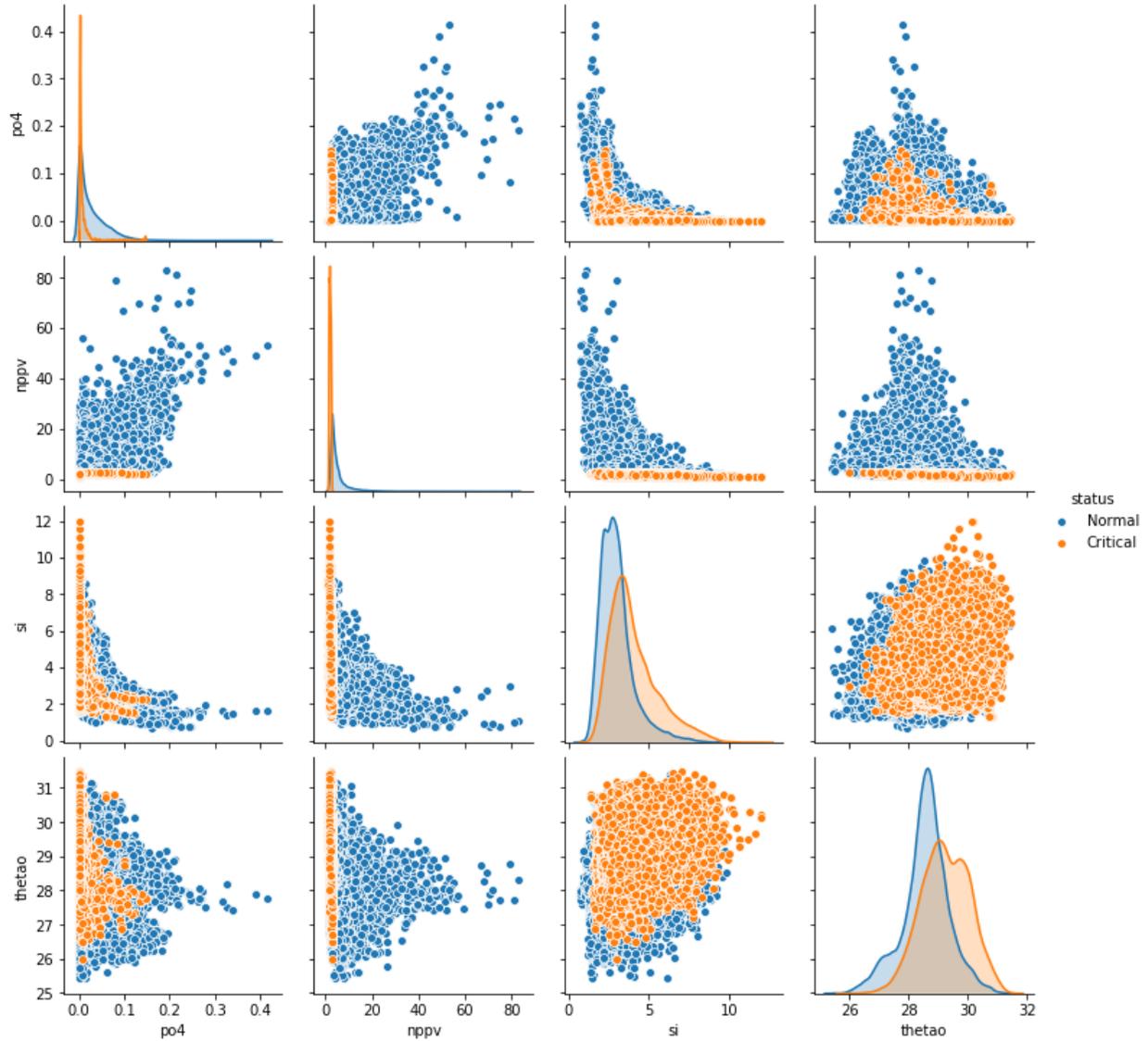

**Fig. 3.** A pair plot for top 4 features which have the highest influence in the classification

The performances of the best four features for classification which are phosphate concentration, the concentration of dissolved silicate, total primary productivity per day and ocean temperature are recorded in table 3. From this, we can see that using these four features, RFC has obtained the best accuracy of 91.51%, followed by CNN, MLP and DTC which obtained 90.12%, 89.59% and 88.86% classification accuracy. We can see that accuracies of all the classifiers have slightly reduced. Looking at the precision scores, both RFC and CNN have equal precision scores of

0.89 followed by MLP and DTC which obtained 0.868 and 0.852 respectively. We can see a drop in precision for all the classifiers but DTC dropped the most which indicated that losing features, it now generating lots of false-positive errors. For recall, we can see that RFC has the highest recall score of 0.884 followed by CNN, MLP and DTC which have recall scores of 0.859, 0.857 and 0.849. We can see that RFC still have decent recall score but all remaining classifiers started to obtain many false-negative errors. Based on accuracy, precision and recall, even after using only 4 features, RFC still proves to be a good classifier for the given problem.

Now for training time, there was a reduction in training time for all the classifiers. However, for testing time, we can see that although RFC, MLP and CNN have retained almost similar testing time, testing time for DTC has increased. For this classification, RFC performs the best among the remaining classifiers.

**Table 3.** Classification performance while using all features

|  | Training Time (ms) | Testing Time (ms) | Accuracy (%) | Precision | Recall |
|---|---|---|---|---|---|
| Random Forest | 5927.74 | 138.989 | 91.51 | 0.89 | 0.884 |
| Decision Tree | 206.08 | 2.346 | 88.86 | 0.852 | 0.849 |
| Multi Layer Perceptron | 8514.07 | 5.186 | 89.59 | 0.868 | 0.857 |
| Convolution Neural Network | 16262.27 | 200.318 | 90.12 | 0.89 | 0.859 |

From these two observations, we can conclude that RFC proves to be the best classifier based on overall performance. But when used only 4 classifiers, RFC's accuracy was approximately equal to DTC built with all features. But in this situation, using DTC with all feature trains around 14 times faster than RFC with 4 features. Therefore, when accuracy is the ultimate goal, RFC could be safely used with all features to obtain an optimum classification accuracy but when resource consumption is concerned, DTC with all features is a much better choice for classification rather than using RFC with lesser features.

## 5 Conclusion

Phytoplanktons are an essential part of the ecosystem and their conservation has a high priority for the sake of maintaining an ecological balance as they are primary producer significantly contributing toward global oxygen production and also as they come under food chain of many marine creatures. They contain chlorophyll which conducts photosynthesis and chlorophylls are visible under near-infrared light bands. For this property, they can be monitored with remote sensing technologies. Remote sensing could also be reanalyzed to estimate many physiochemical features which affect phytoplankton growth.

In this experiment, we have used remote sensing data to establish a classifiable relationship between different physiochemical features to estimate chlorophyll levels based on these features. For our experiment we have constructed time-series data from EU Marine Copernicus Program data and used different classifiers like Random Forest, Decision Tree, etc. and analysed their performances.

From our experiment, we can observe that RFC performs the best for overall performance obtaining classification accuracy of 93.92%, however, its speed is slow. To overcome this, instead of reducing training features, using DTC could provide very fast and reliable performance. DTC obtained 2.4% lower accuracy but trains 19.66 times faster and produces results 83.62 times faster than RFC.

The model could be deployed as an inexpensive global monitoring solution of worldwide chlorophyll levels. This model could be upgraded by training with data from more locations and other features. The extension of the work could be to monitor chlorophyll levels at a much greater depth along with their relation with zooplanktons could be established with the help of remote sensing and learning techniques. Also, their relation with pollutants could be established. All of these together will help toward sustaining an ideal ecosystem.

## 6 Code Availabilty

The code of the project have been made available on GitHub Repository under MIT License of open source distribution of programs and the codes can be freely reused for commercial and non-commercial purpose [19].

## 7 Abbreviations
We have used the following abbreviations throughout the paper.

Table 4. Abbreviations

| Key | Full-Form | Key | Full-Form |
|---|---|---|---|
| DTC | Decision Tree Classifier | nppv | Total Primary Production of Phytoplankton |
| CNN | Convolution Neural Network | Fe | Dissolved Iron |
| RFC | Random Forest Classifier | mlotst | Density Ocean Mixed Layer Thickness |
| MLP | Multi Layer Perceptron | thetao | Temperature |
| spco2 | Surface $CO_2$ Pressure | bottomT | Sea Floor Potential Temperature |
| $O_2$ | Dissolved Oxygen | vo | Northward Velocity |
| Chl | Total Chlorophyll | uo | Eastward velocity |
| $NO_3$ | Dissolved Nitrate | so | Salinity |
| $PO_4$ | Dissolved Phosphate | zos | Sea Surface Height |
| si | Dissolved Silicate | pH | pH |